\documentclass[11pt,twoside]{article}
\usepackage{newpasp}
\usepackage{epsf}

\def\edcomment#1{\iffalse\marginpar{\raggedright\sl#1\/}\else\relax\fi}
\reversemarginpar

\def\ha{\mbox{H$\alpha$}}
\def\nii{\hbox{[N {\sc ii}]}}
\def\sii{\hbox{[S {\sc ii}]}}
\def\ovi{\hbox{O {\sc vi}}}
\def\kms{$\rm {km}~\rm s^{-1}$}

\def\hi{\hbox{H {\sc i}}}

\def\etal{et al.\,}
\def\cmsq{cm$^{-2}$}
\def\cmsqs{cm$^{-2}$ s$^{-1}$}

\begin{document}
\title{Optical Emission from High Velocity Clouds and the 
Ionization Sources in the Galactic Halo}
\author{Benjamin J. Weiner}
\affil{UCO/Lick Observatory, University of California, Santa Cruz, 
Santa Cruz, CA 96064}
\author{Stuart N. Vogel}
\affil{Department of Astronomy, University of Maryland,
College Park, MD 20742}
\author{T.B. Williams}
\affil{Department of Physics \& Astronomy, Rutgers University,
136 Frelinghuysen Rd., Piscataway, NJ 08854}

\begin{abstract}
Optical emission lines have now been detected from about 20
high velocity clouds.  These emission lines -- primarily
\ha, secondarily \nii\ and \sii\ -- are very faint and 
diffuse, spread over the surfaces of the clouds.  
We compile emission line measurements and present a model in which
the \ha\ is recombination caused by photoionizing radiation 
escaping the Milky Way.  In such a model, we infer HVC
distances of 5--30 kpc.  The photoionization
model fails to explain the relatively strong \ha\ emission from 
the Magellanic Stream, and the O VI absorption seen by 
FUSE in HVCs and the MS, which require a second source of
ionization (likely collisional).  Regardless of mechanism, the 
fact that HVCs are
detectable in \ha\ indicates they are not far away enough
to be Local Group objects.  Adopting the HVC distances from 
the model, there appear to be two classes of HVCs:
\ha-bright clouds with low velocity deviations from
Galactic rotation, and often strong \nii, 
which are presumably affiliated with the
Galactic disk; and \ha-faint clouds with high velocity deviations,
which are likely to be infalling gas.
\end{abstract}

\section{Introduction}

High velocity clouds (HVCs) are clouds of neutral hydrogen
which generally appear distinct from the Galactic \hi\ disk.
Cataloged HVCs generally have relatively low \hi\ column densities,
$10^{18} - 10^{20.5}$ \cmsq, and are large, from 0.5 degrees
to tens of degrees across; see the review of Wakker \& van 
Woerden (1997).  No optical counterparts such as stars are
known in HVCs, and the only distance constraints are for a few
HVCs complexes that are seen in absorption against background halo 
stars, at several kpc.  Thus HVC distances, sizes, and masses
are quite uncertain, allowing a wide range of models for
their origin.

Popular models of HVCs include:
recycling of disk gas through a Galactic fountain
({\it e.g.}\ Bregman 1980); stripping from Galactic satellites;
and infall of possibly primordial gas, such as
Local Group models (Giovanelli 1981; Blitz {\it et al.}\ 1999).
These models place HVCs at from $< 10$ kpc to $\sim 1$ Mpc
respectively, a range of 100 in distance and $10^4$ in gas mass.
The Blitz {\it et al.} revival of the Local Group model, incorporating 
dark matter, has attracted recent attention and controversy:
it puts the HVCs at large distances, increasing their masses,
and is potentially related to CDM small halo overpopulation 
issues (Klypin \etal\ 1999; see Gibson \etal, these proceedings).

A wide variety of tests of HVC intragroup models have been proposed
(see Blitz and Gibson \etal, these proceedings).  Observations
of Mg {\sc ii} absorbers and \hi\ surveys of groups have limited the
allowed \hi\ mass in typical groups (Charlton \etal\ 2000,
Zwaan 2001).  However, we still need to understand just what our
HVCs are, even if they are not Local Group objects.

\begin{figure}
\plotfiddle{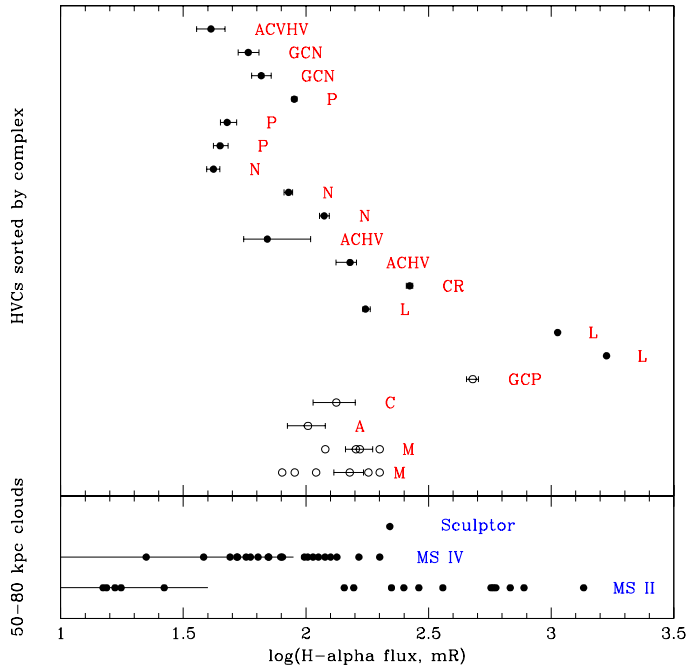}{3.1in}{0}{48}{48}{-150}{-88}
\caption{
\small
H$\alpha$ emission from HVCs (upper panel), and the Magellanic Stream
and Sculptor DSph (lower panel).
The HVCs are grouped by complex (ACVHV, GCN, etc.; Wakker \& van Woerden 1991)
along the y-axis, each line being one HVC.  Open circles:
M, A, and C from Tufte {\it et al.}\ (1998),
and GCP from Bland-Hawthorn {\it et al.}\ (1998).
Filled circles: our data from LCO and CTIO.
}
\end{figure}

Optical emission lines can provide information about
individual HVCs.  Faint,
diffuse optical recombination lines have been seen from
several HVCs (Tufte \etal 1998; Bland-Hawthorn \etal 1998;
Weiner \etal 2000).  The causes of this emission are
not well understood.  It is too faint and diffuse to be
ionization from embedded stars, and is probably 
photoionization, by e.g.\, the Milky Way;
or heating from external sources, such as collisions
with gas in the Galactic halo.
\hi\ HVCs are optically thick to ionizing radiation,
so either photoionization or heating could raise an
ionized skin on the clouds.  Plausible ionization
or heating sources should decline away from the Galaxy,
and optical \ha\ emission can be an indirect indicator
of HVC distance.  Here we describe a sample of HVCs
observed in \ha\ emission and explore the consequences
of a simple ionization and distance model.

\section{Observations}

The only optical emissions detected from HVCs are spatially
diffuse emission lines of \ha, \nii, and \sii.
Several groups have used Fabry-Perot spectroscopy
to detect these emission lines (Weiner \& Williams 1996; 
Tufte \etal\ 1998, 2001; Bland-Hawthorn \etal\ 1998).
These emission lines are much fainter than the sky, 
and ``chopping'' between object and blank sky fields 
is required to achieve sky subtraction to fractions of
a percent (for example spectra see Figure 1 of Weiner \etal\
2000).


\ha\ emission of from 40 to $>1000$ milli-Rayleighs (mR)
has been observed from individual HVCs
(1 Rayleigh = $10^6$ photons cm$^{-2}$ s$^{-1}$ into $4\pi$).
Figure 1 summarizes the results obtained by several groups
and indicates the HVC complexes to which the clouds belong
(Wakker \& van Woerden 1991).  Clouds M, A, and C were
observed by the WHaM group (Tufte \etal 1998), GCP by 
Bland-Hawthorn \etal (1998), and the remainder are our
observations, mostly from Las Campanas using the Dupont 2.5-m
and Wide Field Camera, built by Ray Weymann and collaborators.

A wide range of \ha\ flux is observed -- 1.6 dex among the
various HVCs.
Grossly, clouds in the same complex have similar \ha\ flux.
In a few cases such as the
large northern HVCs M and A, and the intermediate 
velocity complex K, there are measurements
from more than one place on the cloud (Tufte \etal\ 1998;
Haffner \etal\ 2001)
and the \ha\ emission varies by no more than a factor of
2--5.  
In contrast, the emission from the Magellanic Stream is
quite spatially varied, by a factor of 30--40.
\ha\ emission measure is not correlated
with \hi\ column density in the HVC/IVCs (Haffner \etal\ 2001) or the MS.

A few clouds have multiple emission lines measured.
A few of the HVCs brightest in \ha, such as 343+32--140, 
have $\nii/\ha>1$ (see Figure 1 of Weiner \etal 2000).
Similarly high $\nii/\ha$ ratios are observed in the 
extraplanar gas above NGC 891 (Rand 1998).
Some of the fainter HVCs have little or no \nii, possibly
indicating lower metallicity or different ionization state.
In the Magellanic Stream, \nii\ is detectable but
$\nii/\ha \sim 0.2$, a larger decrement than one might
expect just from the metallicity.  This decrement,
combined with the large spatial variations in \ha,
suggests that the MS \ha\ may have a different cause
than the \ha-bright HVCs.

\section{A model of the Galactic ionizing flux}

\begin{figure}
\plotfiddle{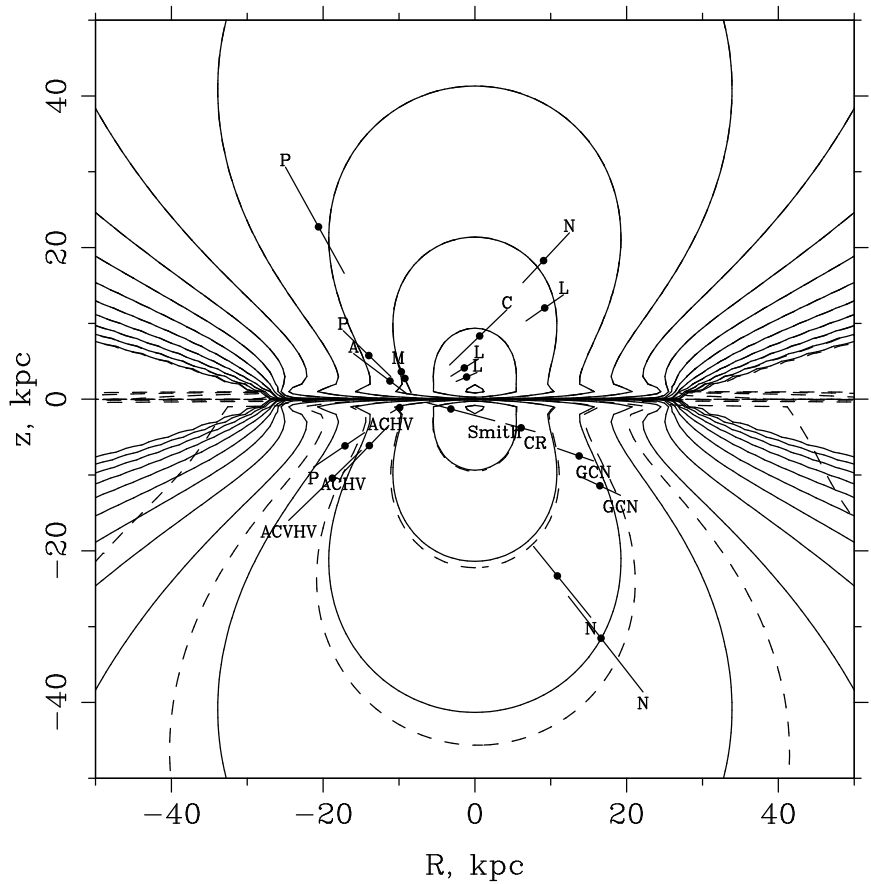}{3.8in}{-90}{67}{67}{-256}{388}
\caption{
\small
A simple model of the ionizing flux emergent from the Galaxy,
with contours of $F_{LC}$ from 1 to $10^{6.5}$ photons cm$^{-2}$ s$^{-1}$ 
by 0.5 dex (dashed contours include the LMC).
Positions of HVCs from the simple photoionization model are indicated
(rotated onto the $\ell= 180\deg - 0\deg$ plane, so that $R$ is
represented accurately).
}
\end{figure}

As a first step to explain the HVC \ha\ emission, 
we consider a simple model: some amount of
Lyman continuum radiation from hot stars escapes the disk
of the Galaxy and ionizes the skin of HVCs.  Since \hi\
HVCs are optically thick to LyC radiation, the recombination
\ha\ emission is proportional to the incident LyC: 0.46
\ha\ photons per Lyc photon at $10^4$ K.  

Figure 2 shows the contours of LyC flux in this model,
using an exponential disk of O stars with total LyC luminosity
$2.7 \times 10^{53}$ photons s$^{-1}$.  We model the Galactic
absorbing layer and Reynolds layer
as a slab with one-sided face-on optical depth
to ionizing photons of $\tau = 2.5$.  The LyC escape fraction,
when averaged over angle, is 2\%.  (See also the model of 
Bland-Hawthorn \& Maloney 1999, and Weiner \etal 2000.)
The small escape fraction implies that galaxies like
the Milky Way do not contribute much to the extragalactic
ionizing background.

We arrived at this model by using the (roughly) known total LyC luminosity,
and tuning $\tau$ to achieve agreement between the known distances and
\ha\ fluxes of HVCs A and M (Danly \etal 1993; Tufte \etal 1998; van
Woerden \etal 1999).  Once we know the distribution of Lyc flux in the
halo, the \ha\ fluxes measured for other HVCs yield their distances,
plotted in Figure 2.  The distance ``error bars'' in Figure 2 assume
a $\pm 50\%$ range in predicted \ha\ due to factors such as cloud
geometry; the observational errors are much smaller.
Variations on the theme of this model produce similar results.  

Figure 3 plots the \ha\ flux of the 20 HVCs
versus their Galactocentric distances as inferred from the model 
(or as measured for A and M). There is a strong correlation, as
may be expected;
the best-fit relation is $F(\ha) \propto D^{-2.2}$, and 
a $-2$ index is a decent fit.  We suggest that there's
nothing particularly special about this model.
Any model (photoionization or not) which 
has a reasonable flux-distance power law index {\it and}
is normalized to the HVCs A and M with known distances
should yield similar predictions for HVC 
distances.\footnote{More complex models, e.g.\,
distributing the O stars in spiral arms rather than a 
smooth disk (Bland-Hawthorn, these proceedings), can make 
a difference in the detailed locations, but for HVCs several 
kpc above the disk, some of the complexity is integrated out.}

\section{The need for another source of ionization}

\begin{figure}
\plotfiddle{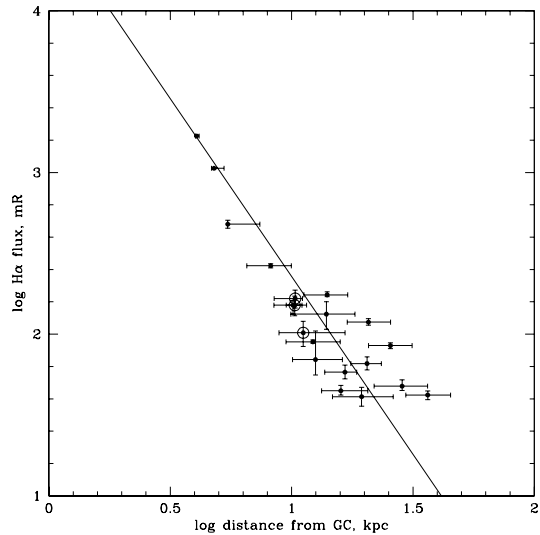}{2.4in}{0}{38}{38}{-120}{-68}
\caption{
\small
For the 20 HVCs from Figure 1, log \ha\ flux versus
distance from the Galactic center, as given by the 
photoionization model of Figure 2.  The circled points
are the HVC complexes A and M used to normalize the model.
The best-fit line has a slope of --2.2.
}
\end{figure}

Unfortunately, photoionization models break down when we look at the 
Magellanic Stream, and another source of ionization is required. 
The MS II cloud is likely at $D \sim 50$ kpc, and
shows $F(\ha)$ from $<40$ to 1300 mR.
From Figure 2, the model LyC flux at MS II (40--50 kpc at $b\sim-90\deg$)
is $\sim 10^5$ photons \cmsqs, predicting $F(\ha) \sim 45$ mR,
30 times lower than the brightest emission and 10 times
below the ``typical'' \ha-bright spots.

There are two problems: the \ha-bright spots are
too bright to be powered by flux escaping from the Milky Way,
and the spatial variation is much larger than expected from 
a photoionized cloud skin.  These problems also exist for MS IV,
and have been known for some time (Weiner \& Williams 1996).
The deficit is
much too large to make up by e.g.\, limb brightening.
Since the model's face-on LyC escape fraction is $e^{-\tau} = 0.08$,
we can't turn it up enough to explain the MS -- most LyC
needs to be absorbed in the Galaxy to power recombination
regions.  Furthermore, photoionization should produce
a nearly uniform \ha\ flux on an \hi\ cloud optically thick
to LyC, yet the MS II emission varies spatially.

Independent evidence for another cause of ionization
comes from FUSE detections of \ovi\ absorption in the MS and in HVCs
(Sembach \etal\ 2000, and these proceedings).  With an
ionization potential of 114 eV, the \ovi\ in these clouds
cannot be produced by photoionization, from Milky Way stars
or otherwise.  Collisional ionization is likely required;
a possible source of heating is interaction between the
\hi\ cloud and hot gas in the halo.
A similar problem is seen in the extraplanar diffuse
ionized gas of NGC 891 and of the Milky Way, where the line 
ratios require something in addition to photoionization (Rand 1998;
Reynolds \etal\ 1999).


These anomalies point to collisional ionization of clouds
in the Galactic halo, presumably through some kind of 
mechanical energy input such as ram pressure or turbulent
mixing, as we previously argued for the Magellanic Stream
(Weiner \& Williams 1996).  But it's difficult to see how 
our 1996 toy
ram-pressure model could produce the peak \ha\ of 1300 mR
now observed in MS II, at least not
from collisions with halo gas of $n \sim 10^{-4}$
cm$^{-3}$.  Perhaps the Stream is colliding with itself
at MS II.  The physics of \ha\ production from 
processes like ram pressure and turbulent mixing is not yet 
well enough understood to make detailed models.  
We need better theories here.

\section{Distances and deviation velocities}

\begin{figure}
\plotfiddle{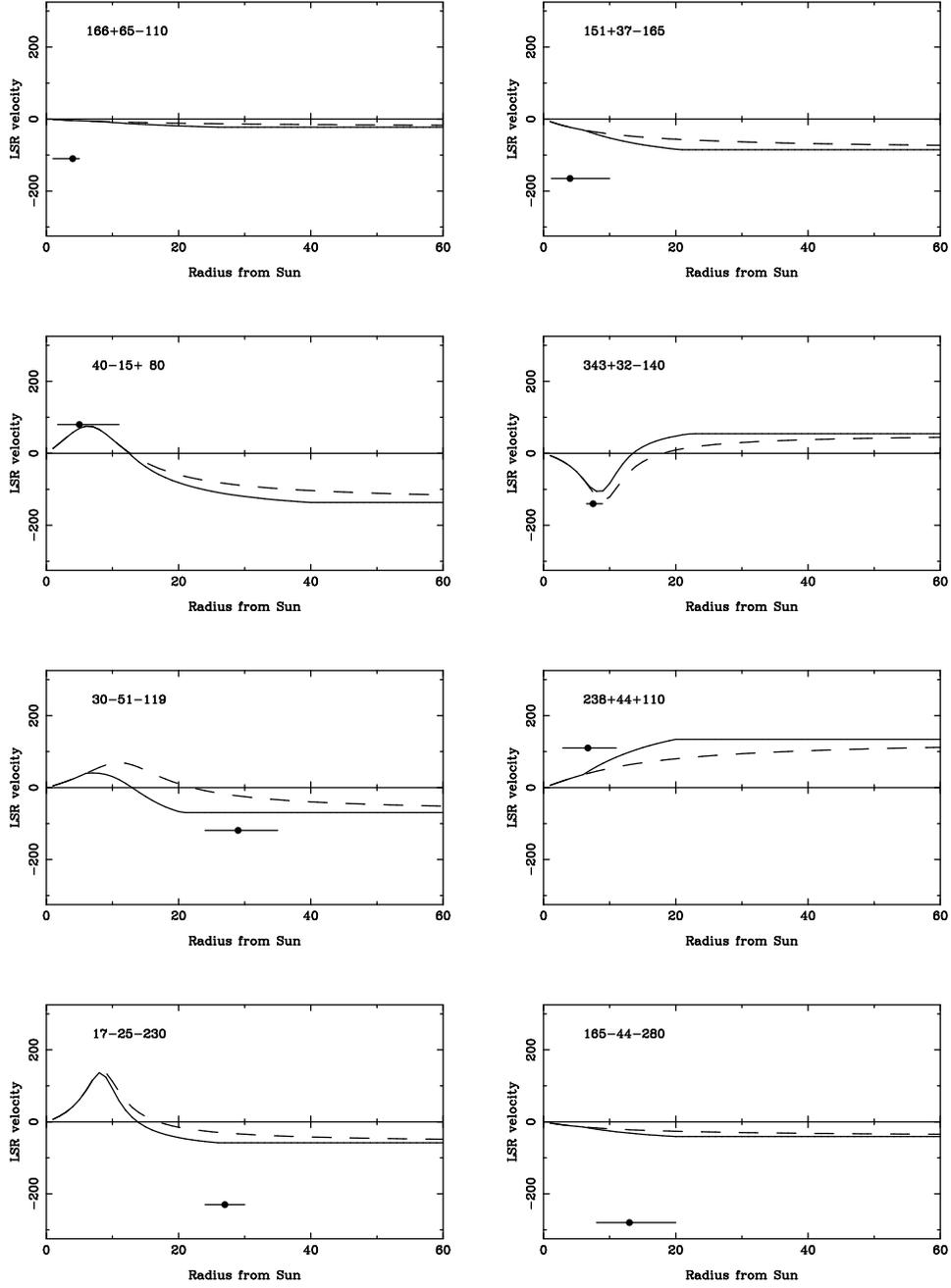}{6.4in}{0}{68}{68}{-206}{-14}
\caption{
\small
For 8 clouds,
HVC LSR velocity and the LSR velocity expected from Galactic 
co-rotation are plotted as a function of distance from the Sun
(dashed line: cylindrical Galactic rotation; solid line:
rotation decaying with $z$ height).  The HVC distance ranges
(points and error bars)
are from the photoionization model of Section 3.
Panels a,b: large complexes M and A.  Panels c,d: \ha-bright
HVCs consistent with Galactic rotation.  Panels e,f: HVCs
with moderate \ha\ and modest deviation velocity.  Panels g,h:
\ha-faint HVCs with large negative deviation velocity.
}
\end{figure}

Despite the uncertainty in the sources of ionization,
it is useful to consider the implications
of the HVC fluxes and distances from the simple model.
There are some reasons to believe that,
for example, photoionization is operating on HVCs A and M
based on the modest range of \ha\ flux
(Tufte \etal 1998).
And even if there is some other ionization mechanism, it ought
to depend on distance from the Milky Way; as we argued above,
models with a fairly generic flux-distance power law will
produce similar distances when normalized to HVCs A and M.

We can use the distance ranges to measure deviation velocities
for the HVC sample.  The deviation
velocity is the difference between the HVC velocity and that
expected for gas in Galactic rotation at the same location
(Wakker \& van Woerden 1991).  It may be more physically
meaningful than the velocity relative to the LSR, but requires
knowing the HVC distance.  

There is some controversy over the most appropriate measure
of HVC velocity.  Blitz \etal\ (1999) showed that the velocity
dispersion of the HVC ensemble is lower in the Galactic
standard of rest (GSR) than in the LSR, and lower yet
in the Local Group standard of rest (LGSR), arguing that HVCs are Local
Group objects.  However, Gibson \etal\ (these proceedings) point 
out that the GSR, LGSR, or any measure of velocity which takes out 
the sinusoid caused by the LSR motion around the Galactic center will 
improve the HVC velocity distribution.

\begin{figure}
\plotfiddle{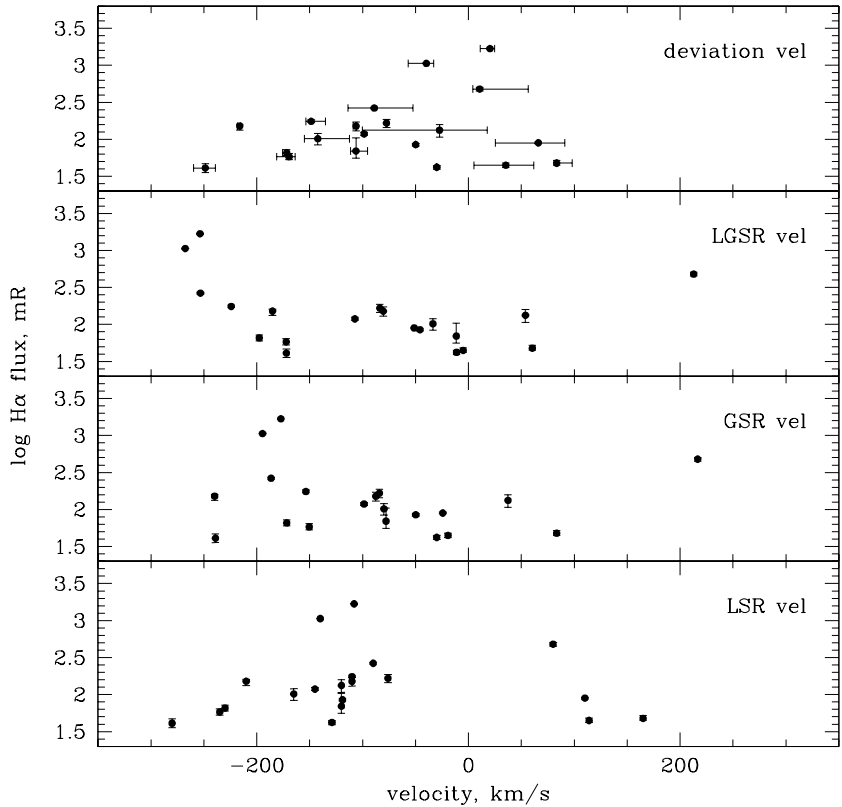}{3.8in}{0}{58}{58}{-176}{-102}
\caption{
\small
For the 20 HVCs from Figure 1, log \ha\ flux versus
various measures of cloud velocity, from top: 
deviation velocity, Local Group (LGSR) frame,
Galactic (GSR) frame, and LSR frame.  By definition,
HVCs must have $|V_{LSR}|>90$ \kms.
}
\end{figure}

Figure 4 shows velocity-distance
plots for 8 of the HVCs in our sample.  The HVC velocity and
distance range are shown together with the expected velocity
for co-rotating gas.  Panels (a,b) show the large, nearby
complexes M and A; since these are toward the Anticenter, the 
co-rotating gas has little radial velocity and the deviation
velocities are fairly large and negative.  Panels (c,d)
show 40--15+80 and 343+32--140.  These clouds are \ha-bright,
and their \ha\ fluxes place them in the inner Galaxy.  At these
distances, they can have deviation velocities of only a few
tens of \kms; they could be disk gas recycled through a
Galactic fountain.  These clouds also have high $\nii/\ha$.
Panels (e,f) show clouds which are 
moderate in \ha\ and deviation velocity.  Panels (g,h)
show 17--25--230 and 165--44--280, very high-velocity clouds
(VHVCs).  These clouds are faint in \ha, putting them at 
tens of kpc distances, and have large deviation velocities 
($\sim -200$ \kms), suggesting infall: their kinetic energy
is too large to have originated in the disk.

In Figure 5, we show the relation between HVC \ha\ flux and 
measures of velocity for the 20 clouds with fluxes.  
These flux-velocity distributions have not been previously accessible 
since there were few \ha\ fluxes or distance estimates.
The sample is not statistically complete and has some minor
selection effects
(e.g.\, the clouds observed from LCO all have $\delta<+10\deg$).
However, we don't expect these to affect the conclusions.

The distribution of flux with LSR velocity (bottom panel of
Figure 5) suggests
that brighter clouds are lower velocity, but LSR 
velocity is less than ideal, since the LSR itself is
moving (and there is an artificial void of clouds at $|V_{LSR}<90|$).
Converting to GSR velocity makes the distribution
somewhat less understandable.  The bright clouds
actually move to higher velocities.  In the LGSR
frame, the same problem occurs to an even greater
degree; the brightest clouds actually have the highest
$|V_{LGSR}|$.  (We assumed an LSR motion of 220 \kms\
toward $(\ell,b) =(90,0)$ in the GSR, and of 280 \kms\ toward
(102,-5) in the LGSR, Einasto \& Lynden-Bell 1982.)
If the advantage of the LGSR frame is that it 
minimizes the velocity dispersion of
HVCs, the highest $|V_{LGSR}|$ are outliers and it is odd
to find the brightest clouds among them, especially since
the brightest clouds also are likely to be the closest 
to the Galaxy.

Using the deviation velocities we found from our HVC
distance model, the flux--velocity distribution suddenly
makes sense (top panel of Figure 5).  
The flux--deviation velocity distribution is compact 
(low dispersion in $V_{dev}$),
peaks near zero $V_{dev}$, and has a tail to negative 
velocities, $V_{dev} \sim -200$ \kms.  The brightest
clouds (such as Figure 4c,4d) have small deviation
velocities -- but because they are co-rotating in the 
inner Galaxy, their GSR velocities are high.  
The tail of \ha-faint
clouds at large negative $V_{dev}$, the VHVCs, are difficult 
to explain with Galactic disk-based models and are 
probably infalling (as was already suggested by Giovanelli 1981;
Wakker \& van Woerden 1991).

\section{Conclusions}

High velocity clouds are routinely (if not easily) detected in \ha.
There are 5 detections in the literature, and
our survey has 15 clouds with good spectra, and has detected \ha\ 
from each; a few other clouds have low-quality spectra and
we can't set any useful limits on the \ha\ flux.  Our faintest
HVC is at 41 mR, while our 2 sigma limit is typically 11 mR,
implying that the sample is not limited by flux.  
Recent results from the WHaM team detect 5 of 6 clouds to equally 
faint limits (Tufte 2001).

The lack of HVCs ultra-faint in \ha\ implies that HVCs are
not ultra-far, for any reasonable ionization mechanism.
In a simple photoionization model normalized to put the
HVCs A and M at the proper distances, the HVCs in our
survey are at 5--30 kpc distances.  One can argue that
another ionization mechanism is operating, as seems to
be the case in the Magellanic Stream.  But even if we
scale the HVCs to the atypical brightest points on the MS,
the HVCs are within $\sim 200$ kpc (and then HVCs A and M
are difficult to explain since they are close).

Adopting the more moderate HVC distances from the simple
photoionization model of Section 3, we can find the amount
by which the clouds deviate from Galactic rotation.
The distribution of \ha\ flux with deviation velocity
is very sensible looking, more so than the distribution 
over $V_{GSR}$ or $V_{LGSR}$.  The clouds brightest in
\ha\ are at low deviation velocities, close to co-rotating.
These clouds also tend to have high $\nii/\ha$ ratios,
and are probably recycled disk gas.  There is a tail
of \ha-faint clouds at high negative $V_{dev}$, and 
nominal tens of kpc distances.  These clouds have too
much kinetic energy to have originated in the disk,
and must be infalling material, perhaps stripped
from dwarf galaxies.  At these distances the infalling clouds are
tidally unstable.

In summary, HVC optical emission fluxes suggest that the
HVCs are not at $\sim 0.5-1$ Mpc, and there are likely two populations:
disk and infall.  The fraction of Lyman continuum photons
escaping from the Milky Way disk is small, $\sim 0.02$.
Several lines of evidence indicate that a second source
beyond photoionization is operating in the Magellanic
Stream and possibly HVCs, and we need better interpretations
of this source and its role in \ha\ production.

\end{document}